# Stabilization and Dissipative Information Transfer of a Superconducting Kerr-Cat Qubit

U. KORKMAZ and D. TÜRKPENÇE

*Abstract*— Today, the competition to build a quantum computer continues, and the number of qubits in hardware is increasing rapidly. However, the quantum noise that comes with this process reduces the performance of algorithmic applications, so alternative ways in quantum computer architecture and implementation of algorithms are discussed on the one hand. One of these alternative ways is the hybridization of the circuit-based quantum computing model with the dissipative-based computing model. Here, the goal is to apply the part of the algorithm that provides the quantum advantage with the quantum circuit model, and the remaining part with the dissipative model, which is less affected by noise. This scheme is of importance to quantum machine learning algorithms that involve highly repetitive processes and are thus susceptible to noise. In this study, we examine dissipative information transfer to a qubit model called Cat-Qubit. This model is especially important for the dissipative-based version of the binary quantum classification, which is the basic processing unit of quantum machine learning algorithms. On the other hand, Cat-Qubit architecture, which has the potential to easily implement activation-like functions in artificial neural networks due to its rich physics, also offers an alternative hardware opportunity for quantum artificial neural networks. Numerical calculations exhibit successful transfer of quantum information from reservoir qubits by a repeated-interactions-based dissipative scheme.

## I. INTRODUCTION

THE ERA in which today's computers cope with noise is called the noisy intermediate scale quantum (NISQ) era [1], and various types of research on noise suppression have shown up. One of the most prominent of these searches is that the part of the calculation that provides a quantum advantage is done together with the ordinary quantum circuit model, and classical computers do the part that does not provide a quantum advantage. Although this model is not a complete quantum computing model, there is a quantum acceleration in the process. That's why this model is called quantum acceleration. In our previous work, we presented a dissipation-based quantum classifier model [2], [3].

Before starting the subject, let's briefly talk about the harmonic oscillator, creator and annihilator processors, which have a very important place in the subject. The quantum Hamiltonian of a simple one-dimensional harmonic oscillator with mass $m$ and frequency $\omega_m$ is expressed as:

$$H = \frac{p^2}{2m} + \frac{1}{2}m\omega_m^2 q^2 \qquad (1)$$

Here, $p$ and $q$ are position and momentum operators, respectively, and as it is known, $[q,p] = i\hbar$ provides the commutation relation. The reduced Planck's constant, $\hbar$, is used here.

The operators that provide the commutation relation $[\alpha,\alpha^\dagger] = 1$ under the $q = q_0(\alpha^\dagger + \alpha)$ and $p = p_0(\alpha^\dagger - \alpha)$ transformations are called the annihilation $\alpha$ and the creation operator $\alpha^\dagger$. Here $q_0 = \sqrt{\hbar/2m\omega_m}$ ; $p_0 = \sqrt{\hbar m\omega_m/2}$. When we replace these transformations and the operators that provide the commutation relation in Eq. (1), it turns into

$$H = \hbar\omega_m\left(\alpha^\dagger\alpha + \frac{1}{2}\right). \qquad (2)$$

The $\alpha^\dagger\alpha$ operator is called the number operator because it measures the number of quanta covered by the oscillator. The eigenstates of the number operator are $n = 0, 1, 2, \cdots$ and $|n\rangle$ number states, which correspond to the presence of exactly $n$ quanta in the oscillator:

$$\alpha^\dagger\alpha|n\rangle = n|n\rangle. \qquad (3)$$

It follows from this that $|n\rangle$ is also the eigenstates of the Hamiltonian of Eq. (2), so we can express the eigenvalues as:

$$E_n = \hbar\omega_m\left(n + \frac{1}{2}\right) \qquad (4)$$

It can be demonstrated that the creation and annihilation operators, respectively, increase and reduce the number of quanta in a number state:

$$\alpha|n\rangle = \sqrt{n}|n-1\rangle, \quad \alpha^\dagger|n\rangle = \sqrt{n+1}|n+1\rangle. \qquad (5)$$

"Schrödinger-Cat" states are known as macroscopically recognizable overlapping states of a single-mode quantized electromagnetic field, and in the field of quantum optics, work on the creation of these states and the study of their properties has recently increased. According to the interpretation that is most frequently used in the literature, a macroscopic superposition state (MSS) is a superposition of ordinary coherent states with 180°-differences in their relative amplitudes [4], [5]. In other words, $|\alpha\rangle$ is the coherent state of the unimodal field, where $|\alpha\rangle$ is complex, and $\alpha|\alpha\rangle = \alpha|\alpha\rangle$,

**UFUK KORKMAZ**, is with Department of Electrical Engineering, Istanbul Technical University, Istanbul, Turkey, (e-mail: ufukkorkmaz@itu.edu.tr).

https://orcid.org/0000-0001-5836-5262

**DENİZ TÜRKPENÇE**, is with Department of Electrical Engineering, Istanbul Technical University, Istanbul, Turkey, (e-mail: dturkpence@itu.edu.tr).

https://orcid.org/ 0000-0002-5182-374X

where $\mathfrak{a}$ is the mod annihilation operator. As a result, $|-\alpha\rangle$ is likewise a coherent field state with $\mathfrak{a}|-\alpha\rangle = -\alpha|-\alpha\rangle$. Superpositions are used to describe Schrödinger-Cat states (or MSSs). In contrast to parametric amplification, a process that uses energy can produce a pure state using a limited amount of energy. Schrödinger-Cat states, which are quantum mechanical superpositions of two coherent out-of-phase states, are these pure states [6]. The superposition states of Schrödinger-Cat states (or MSSs) are expressed as follows:

$$|\psi\rangle = c_+|\alpha\rangle + c_-|-\alpha\rangle \quad (6)$$

Such cases are eigenstates of $\mathfrak{a}^2$ such that $\mathfrak{a}^2|\psi\rangle = \alpha^2|\psi\rangle$. The remarkable situations called as the even and odd parity coherent cases (Eqs. (7.a) and (7.b)), respectively, were of great interest.

$$|C_\alpha^+\rangle = \mathcal{N}^+(|\alpha\rangle + |-\alpha\rangle) \quad (7.a)$$
$$|C_\alpha^-\rangle = \mathcal{N}^-(|\alpha\rangle - |-\alpha\rangle) \quad (7.b)$$

Here $\mathcal{N}^\pm = 1/\sqrt{2(1 \pm e^{-2|\alpha|^2})}$. There is $\bar{n} = |\alpha|^2$ relation between $\alpha$ and average photon number [7], [8].

Logical states in quantum information theory can be represented as two orthogonal pure states, most easily via a single two-level system. Therefore, logical 0 and 1, respectively, can be encoded in the excited state $|e\rangle$ and ground state $|g\rangle$. Qubits, as opposed to just bits, are the name given to logical states since they have the ability to combine coherently [9]. It powers quantum computing thanks to its ability to treat the coherent superposition of large sets of physical systems as units. One need not limit oneself to two-dimensional Hilbert space-codifying physical systems. It is more useful for some cases to code the logical states as a superposition spanning many base cases. An unwelcome and perhaps out-of-control single interaction in the encoded information takes place when a device that permits qubit encoding is connected to a disturbed environment, and this interaction may appear as an error. The type of logical mistake is determined by binding to the logical foundation. Although the binding to the medium is fixed, we have freedom in how we encode the qubits, therefore the foundation for logical encoding can affect the sort of mistake that happens [9].

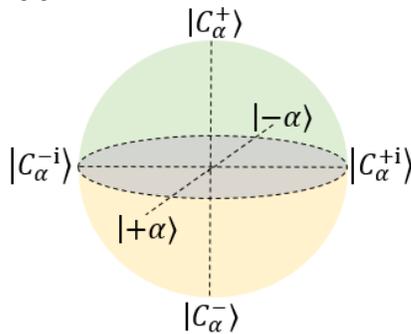

Fig.1. Bloch ball representation of a Cat qubit. Basis state in six different points are labelled.

Schematic of the Cat qubit Bloch representation is shown in Fig. 1. It is simple to confirm that the even-parity Cat state ($|C_\alpha^+\rangle$) contains only even-energy eigenstates and the odd-parity Cat state ($|C_\alpha^-\rangle$) contains only odd-energy eigenstates, regardless of $\alpha$ [9]. The logical encoding can be represented as follows for a single qubit due to the orthogonality of the two states.

$$|C_\alpha^+\rangle = |\bar{0}\rangle \quad (8.a)$$
$$|C_\alpha^-\rangle = |\bar{1}\rangle \quad (8.b)$$

A higher-dimensional system's two-dimensional subspace call a logical qubit has been chosen to enable the identification and rectification of specific errors. It frequently takes arbitrary and exact control over the whole system to manipulate encoded information. No matter if logical operations are based on huge dimensional modes like oscillators or multiple physical qubits, constituent parts in actual devices will always have residual interactions that must be taken into consideration when creating logical operations [7], [10].

The potential applications of the Cat-qubit structure are described more clearly in the following statements. These non-classical quantum harmonic oscillator states can be utilized as a system for weak force detection [11], quantum information processing [12], and quantum repeaters [13]. Recently, a study on flying Cat-qubits, which have the potential for use in quantum communication, has also been published [14]. Once the potential implications were expressed in the literature, the feasibility of cat qubits for actual physical systems became significant. Therefore, stabilization and operational proposals have appeared in the literature [15]–[18]. Moreover, advanced protocols including error correction schemes were also reported [19]–[21].

The Kerr-Cat qubit, like the superconducting transmon qubit, is considered bi-level [15], [21]. In the transient dynamics of the Kerr-non-linear resonators (KNR), the Cat states are obtained by connecting a superconducting qubit to a microwave resonator [8]. These obtained states are of great importance in understanding the mismatch in precision measurements in macroscopic systems, which are very important in quantum computing. These states, however, are challenging to produce and manage with high precision because to their vulnerability to undesired interactions and photon loss. To do this, we develop a collisional quantum master equation based on a series of recurrent encounters. Particularly when reservoir states must be clearly stated, as in our current issue, collision models have shown to be very helpful in understanding open quantum dynamics [22]–[26]. Because of the additivity and divisibility qualities of quantum dynamic maps, collision dynamics may yield a stable state given any amount of input data and without the requirement for time-dependent control [27]–[29].

Designing qubits based on atomic structure or superconducting architecture rather than using relatively less stable cat-qubits may be seen as a pitfall that worsens the process. However, as shown by the exciting potential for applications in quantum communication and detection, working on this model is worth it. On the other hand, the ability to be applied by cavity-quantum electrodynamics systems with extensive quantum control experience and network formation capability makes this model suitable as a machine learning architecture.

Currently, quantum machine learning is generally discussed based on the standard quantum circuit model. Artificial neural networks that play an important role in machine learning and the binary classifier, which is the basic processing unit, are simulated using quantum software based on the quantum circuit [30]–[32]. On the other hand, today's quantum computers are susceptible to noise, so the number of qubits that will provide accurate results is limited. Therefore, noise-resistant algorithms or hardware-efficient solutions are being emphasized [33], [34].

One of these solutions is the dissipation-based quantum computing model that operates with an open quantum system model [35]. It is known that this model type is more advantageous against noise effects. For example, we recently published a study on how a binary classifier quantum perceptron would work according to this model [36]. According to this study, a probe qubit is an open quantum system connected to independent information reservoirs, and the classification result is encoded as a stationary state magnetization. In our current study, where the probe qubit can be described as a cat-qubit, we are numerically testing the stabilization of the cat-qubit and the transfer of information from the information reservoir to the cat-qubit.

In this study, we show that a quantum information classifier can be successfully created with the dissipation-based protocol. A different version of the quantum acceleration model is to use a dissipation-based protocol instead of using a classical computer in the model. Thus, the propose dissipation-based model will both cause less noise and eliminate the difficulty of a classical computer dealing with quantum data.

We numerically examine a dissipation-based information transfer process that would lead to a Cat Qubit-based binary classifier model. Cat-Qubit is an alternative physical qubit model that creates logical bases with appropriate superpositions of so-called coherent states of quantum harmonic oscillator [4], [5], [7], [8]. This physical model, in which the control parameters are richer, has the potential to be evaluated as a quantum neuron, thanks to Kerr non-linearity-based activation possibility. The dissipation model in our study is described by the collision model, which is called the collision model and is based on repeated interactions.

## II. MODEL AND SYSTEM DYNAMICS

### A. Kerr-Cat Qubit Hamiltonian

Some protected qubits, such as these bi-level qubits (a type of bosonic qubit) mentioned in the previous section, are stabilized by a Hamiltonian [21]. Therefore, the following Hamiltonian is used to stabilize the Kerr-Cat qubit.

$$H = \hbar(K\hat{a}^{\dagger 2}\hat{a}^2 - \epsilon_2(\hat{a}^{\dagger 2} + \hat{a}^2) + \Delta_{ar}\hat{a}^\dagger\hat{a}) \quad (9)$$

Here, K is the Kerr nonlinearity, $\epsilon_2$ is the squeezing-drive strength, $\Delta_{ar}: \omega_0 - \omega_r$ is the frequency with respect to the rotary coordinate system where $\omega_0 = 37.7\ GHZ$ is the cavity frequency and $\omega_r$ is the resonator frequency, respectively. $\hbar = 1$ throughout the entire study. This interpretation of the Hamiltonian demonstrates that the two stables $|\pm\alpha\rangle$ states and $\alpha = (\epsilon_2/K)^{1/2}$, the eigenstates of the annihilation operator $\hat{a}$, are also degenerate eigenstates of Eq. (9) with energy $|\epsilon_2|^2/K$. Equivalently, the even-odd parity states of $|C_\alpha^\pm\rangle$ are the eigenstates of $H$ [8], [15].

### B. Open Quantum System

A subset of supervised learning problems called binary classification tries to create a classification rule that depicts a mapping from $\mathcal{X} \to \{0,1\}$ abstract feature spaces to a collection of binary labels. Here, $\mathcal{X}$ denotes the $\mathcal{D} = (\mathcal{X}, \mathcal{Y})$ training set's input data field. The binary labels' associated output data field is $\mathcal{Y}$. Another way to state the training set is as follows:

$$(\mathcal{X}, \mathcal{Y}) = \{(x_i, y_i)\}^M \quad (10)$$

These pairs in relation to the $M$ elements, $x_i$ is a vector representing the unclassified input data, while $y_i$ represents the output vector corresponding to $x_i$. Because the input feature data is a vector in N-dimensional real space, the $\mathbb{R}^N \to \{0,1\}$ map is defined by the binary classification. The perceptron, the primary processing unit of artificial neural networks, is the most basic binary classifier and it is connected with mathematical models [37]. The binary classification model, inspired by a biological neuron, $x = [x_1, \cdots x_N]^T$ is defined as $z = f(x^T w)$ in terms of these vectors representing the input information with the corresponding vector and the weight vector $w = [w_1, \cdots w_N]^T$. Here $f$ is the linear or non-linear activation function, which has an important role in determining the classification rule. The classification is performed by the model as $z \equiv 1$ if $z = f(x^T w)$, not $z \equiv 0$.

A weighted combination of input data samples can be used with the conventional binary classification model.

$$x^T w = \sum_i w_i x_i \quad (11)$$

We may construct our model using linear combinations of quantum dynamic maps, each of which represents the attachment of the system to a separate reservoir, as motivated by this classical point of view.

$$\Lambda_t[\varrho_0] = \sum_i P_i \Phi_t^{(i)}[\varrho_0] \quad (12)$$

In this case, $\varrho_0$ stands for the probe qubit from which the output data is read, and $P_i$ is a positive value that describes the likelihood of the system encountering each reservoir. We imply that every map in Eq. (13) results in a steady state.

$$\Phi_t^{(i)}[\varrho_0] = Tr_{\mathcal{R}_i}[\Lambda_t(\varrho_0 \otimes \varrho_{\mathcal{R}_i})\Lambda_t^\dagger] \quad (13)$$

$Tr_i$ is partial trace on unit $ith$. $\varrho_{\mathcal{R}_i}$ is the quantum state of the $ith$ reservoir. The completely positive trace-preserving quantum dynamical map (CPTP) [24] represented as $\Lambda_t[\varrho] = -i[H_t, \varrho] + \kappa_1 \mathcal{L}_0[\hat{a}] + \kappa_2 \mathcal{L}_0[\hat{a}^2]$ serves as a defining characteristic of the system. Also, a dynamic map that provides $\Phi_{t+s} = \Phi_t(\Phi_s[\varrho])$ complete positivity (CP) for $t$ and $s \geq 0$ is called a CP divisible map. $H_t = K\hat{a}^{\dagger 2}\hat{a}^2 - \epsilon_2(\hat{a}^{\dagger 2} + \hat{a}^2) + \Delta_{ar}\hat{a}^\dagger\hat{a}$ is the Hamiltonian that represents a consistent drive on the Kerr-Cat qubit. $\mathcal{L}[o] \equiv 2o\varrho o^\dagger - o^\dagger\varrho o - \varrho oo^\dagger$ is the standard Lindblad super operator. The two super operator constants $\kappa_1$ and $\kappa_2$ describe the losses of one photon and two photons. $\kappa_1 = 1/T_1$ [8]. Eq. (13) is the explicit quantum equivalent of Eq. (12) and is only valid if it satisfies the addition and divisibility conditions of quantum dynamic maps [27], [29], [38]. According to reports, when the weak coupling condition is satisfied, both the complete positivity and the divisibility requirement for quantum dynamic maps hold true [29]. We must focus on two key points in order to comprehend our model more fully. The suggested classifier's dynamics are first presented, which is a collision model in which a probe qubit connects sequentially with identical and non-interacting reservoir units with a limited time $\tau$. Second, we refer to these reservoirs as information reservoirs as they are made up of identical qubits that have been idealized using particular collision model-derived characteristics. These formulations read $\varrho(0) = \varrho_S(0) \otimes \varrho_{\mathcal{R}_i}$ as the initial quantum state of the entire system. Each information reservoir is made up of a set of identical, non-interacting qubits in pure quantum states, as was already mentioned.

$$\varrho_{\mathcal{R}_i} = \otimes_{k=1}^{n} \varrho_k(\theta_i, \phi_i) \quad (14)$$

Here, $\varrho_k(\theta_i, \phi_i)$ is the $kth$ subunit of the $ith$ information reservoir. The Hamiltonian of the system and the $ith$ information reservoir is expressed as:

$$H_0 = \hbar \left( K a_0^{\dagger 2} a_0^2 - \epsilon_2 (a_0^{\dagger 2} + a_0^2) + \Delta_{ar} a_0^\dagger a_0 + \Delta_{ir} \sum_{i=1}^{n} a_{k_i}^\dagger a_{k_i} \right) \quad (15)$$

and the interaction term is expressed as:

$$H_{int} = \hbar \epsilon_x \sum_{i=1}^{n} \left( |C_\alpha^+\rangle_0 \langle C_\alpha^-| \otimes |C_\alpha^-\rangle_1 \langle C_\alpha^+| + H.c. \right) \quad (16)$$

where $\hbar$ is the reduced Planck constant, $\Delta_{ar} = \Delta_{ir}$ and $\epsilon_x$ is the Cat-Rabi driver constant. Rearranging Eq. (13) as cascading partial trace operations to describe dynamic maps via the collision model, we can write it as follows:

$$\Phi_{n\tau}^{(i)}[\varrho_0] = Tr_n \left[ \Lambda_{0i_n} \cdots Tr_1 \left[ \Lambda_{0i_1} (\varrho_0 \otimes \varrho_{\mathcal{R}_{i_1}}) \Lambda_{0i_1}^\dagger \right] \otimes \cdots \otimes \varrho_{\mathcal{R}_{i_n}} \Lambda_{0i_n}^\dagger \right] \quad (17)$$

Here, $\Lambda_{0i}[\varrho] = -i[H, \varrho] + \kappa_1 \mathcal{L}_0[a] + \kappa_2 \mathcal{L}_0[a^2]$ is a dynamic map that provides the divisibility of CP where $H = H_0 + H_{int}$ and $\varrho = \varrho_0 \otimes \varrho_{\mathcal{R}_{i_n}}$. Here, $n\tau$ is the amount of time required for n collisions, and $\Lambda_{0i}$ is the non-unitary super operator acting on both the system and reservoir qubits. $n$ is often a finite number that is big enough to stabilize the probe qubit. In the Markov technique, the system state equals that of one of the reservoir units after a sufficient number of interactions. In other words, it is by this discrete dynamic development that the system temperatures get to the thermal reservoir temperature. Quantum homogenization is the name of this procedure [22]. As a result, we simulate our system numerically, which is explained by the following master equation. This process was performed with the Qutip toolbox [39].

$$\dot{\varrho}_0 = -i[H, \varrho] + \kappa_1 \mathcal{L}_0[a] + \kappa_2 \mathcal{L}_0[a^2] \quad (18)$$

### III. NUMERICAL SIMULATION RESULTS

For numerical computations we use the parameters of superconducting circuits, which has become one of the most successful platforms for quantum information processing applications [40]–[42]. Transmon qubits can be coupled via a resonator bus, where interactions are mediated by the exchange of virtual photons [43]. In this architecture, the coupling forces between the qubits can be controlled through the coupling distributed to the transmission line resonator. A superconducting circuit with weakly coupled transmon qubits typically has resonator frequency $w_r \sim 1 - 10 \, GHz$ and qubit-resonator coupling $g \sim 1 - 500 \, MHz$ and effective qubit-qubit coupling $J \sim 1 - 100 \, MHz$ and qubit energy relaxation time $T_1 \sim 20 - 60 \, \mu s$ [42], [43].

Fig. 2 shows the variation of the Kerr-Cat qubit populations and magnetization depending on the system parameters. Here our qubit is unstable.

It is seen in Fig. 3 that the oscillation in population states decreases depending on the system parameters. In Figs. 2 and 3, $T_1 = 15.5 \times 10^{-6}$ s and $\tau = 3 \times 10^{-9} s$ are scaled with $w_r = 2\pi \times 10^9 Hz$. Accordingly, it is scaled as $T_1 w_r = 558 \times 10^3$. $\tau_{sc} \cong 113$ corresponds to $\frac{T_1 w_r}{\tau_{sc}} \cong 5000$ collisions. Thus, 100000 collisions equal $20 T_1$ times, approximately $30 \times 10^{-5} s$. This indicates that the qubit remains stable for long enough.

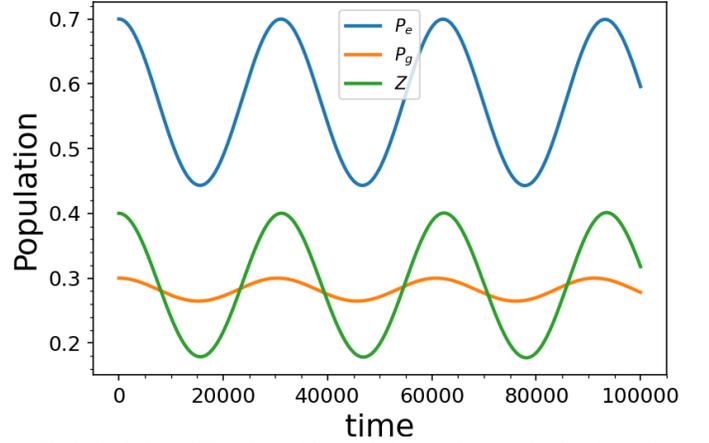

Fig.2. Variation of Kerr-Cat qubit populations and magnetization depending on system parameters. Here, $P_e$ is the excited state population, $P_g$ is the ground state population, and $Z = P_e - P_g$ magnetization. $K = 1.12 \times 10^{-6}$, $\epsilon_2 = 2.25 \times 10^{-6}$, $\kappa_1 = 1.71 \times 10^{-6}$, $\kappa_2 = 3.34 \times 10^{-5}$ and $\Delta_{ar} = 1.00 \times 10^{-4}$ are dimensionless and scale with $w_r$.

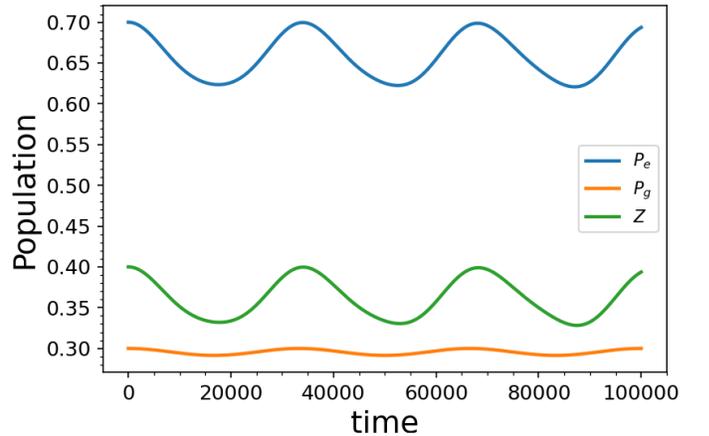

Fig.3. Variation of Kerr-Cat qubit populations and magnetization depending on system parameters. Here, $P_e$ is the excited state population, $P_g$ is the ground state population, and $Z = P_e - P_g$ magnetization. $K = 1.12 \times 10^{-6}$, $\epsilon_2 = 2.25 \times 10^{-5}$, $\kappa_1 = 1.71 \times 10^{-6}$, $\kappa_2 = 3.34 \times 10^{-5}$ and $\Delta_{ar} = 1.00 \times 10^{-4}$ are dimensionless and scale with $w_r$.

In Fig. 4, it is seen that our qubit has become stable depending on the system parameters. After stabilizing our qubit, we will use the time dependent Hamiltonian to examine the activation states.

$$H(t) = K \hat{a}^{\dagger 2} \hat{a}^2 - \epsilon_2(t)(\hat{a}^{\dagger 2} + \hat{a}^2) + \Delta_{ar} \hat{a}^\dagger \hat{a} \quad (19)$$

Here, there are $t \gg \tau$ and $\tau K = 5$ relations for $\epsilon_2(t) = \epsilon_2^0[1 - exp(-t^4/\tau^4)]$, $\epsilon_2(t = 0) = 0$ and $\epsilon_2(t) \sim \epsilon_2^0 = 4K$.

TABLE I
FIGS. 1, 2, and 3 RESULTS

|  | $w_r$, [GHz] | $K/w_r$ | $\epsilon_2/w_r$ | $\kappa_1/w_r$ | $\kappa_2/w_r$ | $\Delta_{ar}/w_r$ |
|---|---|---|---|---|---|---|
| Fig. 2 | $2\pi$ | $1.12 \times 10^{-6}$ | $2.25 \times 10^{-6}$ | $1.71 \times 10^{-6}$ | $3.34 \times 10^{-5}$ | $1.00 \times 10^{-4}$ |
| Fig. 3 | $2\pi$ | $1.12 \times 10^{-6}$ | $2.25 \times 10^{-5}$ | $1.71 \times 10^{-6}$ | $3.34 \times 10^{-5}$ | $1.00 \times 10^{-4}$ |
| Fig. 4 | $2\pi$ | $1.12 \times 10^{-6}$ | $2.25 \times 10^{-6}$ | $1.71 \times 10^{-6}$, | $3.34 \times 10^{-6}$ | $5.80 \times 10^{-6}$ |

In Fig. 5, we consider a single information reservoir $\varrho_1 = |\psi(\theta_1, \phi_1)\rangle\langle\psi(\theta_1, \phi_1)|$ in contact with a non-decay probe Kerr-Cat qubit. Bloch vector was expressed using the $|\psi(\theta_1, \phi_1)\rangle = \cos\frac{\theta_1}{2}|C_\alpha^+\rangle + \sin\frac{\theta_1}{2}e^{-i\phi_1}|C_\alpha^-\rangle$ Kerr-Cat qubit. While following the amplitude parameter for the classification result in our proposal, we monitor the equilibrium dynamics over a steady state. In this simplest case, stationary dynamics yields no classification results, but provides an instructive example of quantum homogenization. As shown in Fig. 5 (a) and (b), the probe qubit magnetization converges to the magnetization of the reservoir units with the amplitude parameters $\theta_1 = 0$ and $\theta_1 = \pi$, respectively. The probe prepares as the Kerr-Cat qubit $\varrho_0 = |+\rangle\langle+|$, which is originally intended to provide a null magnetization. Taken in $|+\rangle = |+\alpha\rangle = 2^{-1/2}[(|C_\alpha^+\rangle + |C_\alpha^-\rangle) + (|C_\alpha^+\rangle - |C_\alpha^-\rangle)]$ form. Without losing generalization, we get $\phi_i = 0$ in our calculations relative to $\theta$.

The smooth and monotonic convergence of the equilibrium curves exhibits a Markov evolution demonstrating the success of CP divisibility collision dynamics. Here homogenization means a equilibration process in which the quantum state of the system becomes the same as that of the reservoir density matrix with its diagonal inputs [23], [24].

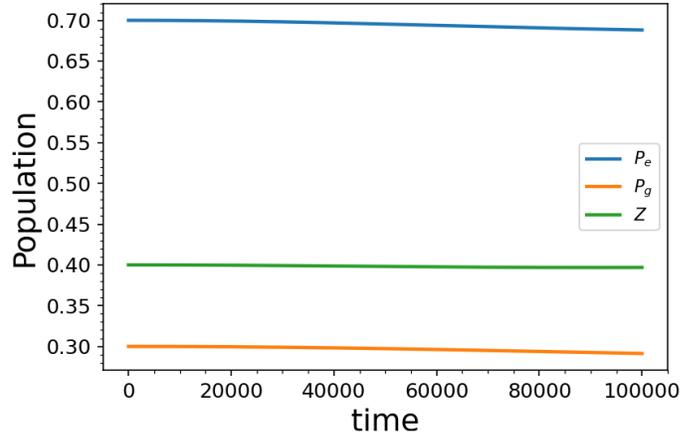

Fig.4. Variation of Kerr-Cat qubit populations and magnetization depending on system parameters. Here, $P_e$ is the excited state population, $P_g$ is the ground state population, and $Z = P_e - P_g$ magnetization. $K = 1.12 \times 10^{-6}$, $\epsilon_2 = 2.25 \times 10^{-6}$, $\kappa_1 = 1.71 \times 10^{-6}$, $\kappa_2 = 3.34 \times 10^{-6}$ and $\Delta_{ar} = 5.80 \times 10^{-6}$ are dimensionless and scale with $w_r$.

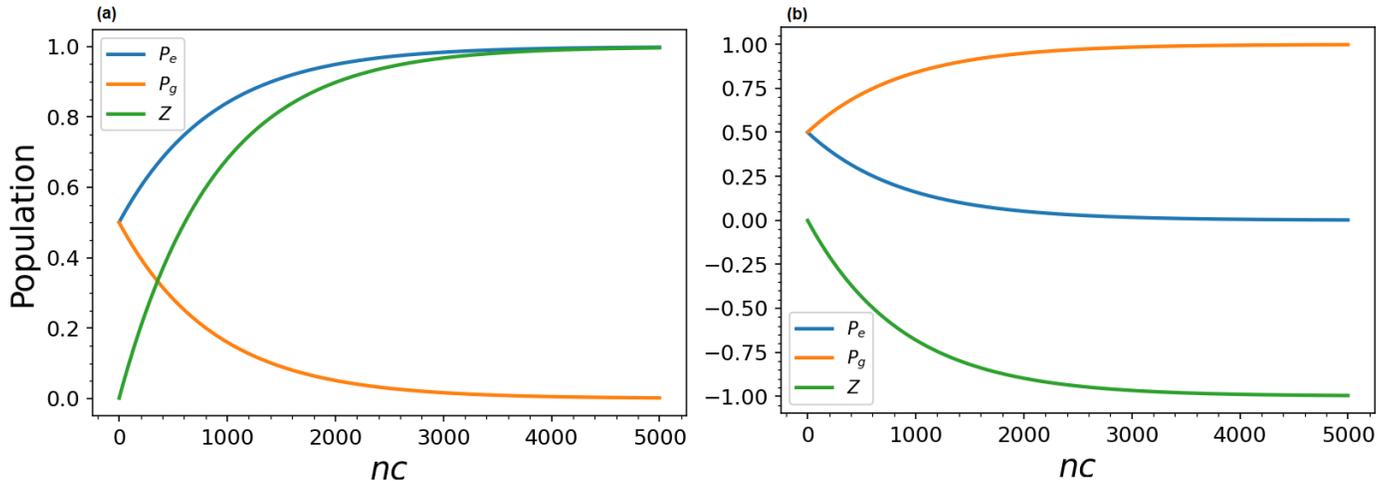

Fig. 5. Variation of Kerr-Cat qubit populations and magnetization depending on system parameters. Here, $P_e$ is the excited state population, $P_g$ is the ground state population, and $Z = P_e - P_g$ magnetization. $K = 1.12 \times 10^{-6}$, $\epsilon_2 = 2.70 \times 10^{-4}$, $\kappa_1 = 1.71 \times 10^{-6}$, $\kappa_2 = 3.34 \times 10^{-4}$, $\Delta_{as} = 5.80 \times 10^{-6}$ and $\tau = 113.01$ are dimensionless and scale with $w_r$.

As the amplitude parameter of the probe qubit is chosen as the identifying merit of the classification, we introduce the classification rule using the probe qubit's $Z = Tr[\varrho\sigma_z]$ steady state magnetization where $\sigma_z$ is the Pauli-z operator. Hence, the binary decision of the classifier in steady state with binary labels 0 and 1 reads

$$Decision \begin{cases} 0, & Z \geq 0 \\ 1, & otherwise \end{cases}. \quad (20)$$

The above-mentioned work demonstrates the possibility of a dissipative information transfer process for the cat-qubit system using realistic parameters. In this application, a single information reservoir is repeatedly interacted with a probe cat-qubit system, resulting in open quantum system evolution. The probe system responds with a stable magnetization value in its stationary state. Since the quantum reservoir input represents quantum information, this demonstration is, in fact, the simplest binary classification example for a single input quantum information, according to the classification rule expressed in the text. Eq. (12), expressed in terms of dynamical maps, shows that this process can be easily generalized to multiple quantum information input situations. The convex combination additivity of the density matrix formulation representing open quantum systems suggests that the example in this study, which responds in binary classification form with a single input, has the potential to be a general open quantum classifier for cat-qubits.

## IV. CONCLUSION

In this study, we investigate whether a quantum neuron model with dissipative information transfer is possible with numerical methods. For this, exact diagonalization methods are used with realistic parameters in the Qutip toolbox. First, as can be seen in Figs. 2, 3, and 4 the stabilization of the Cat-Kerr qubit, which takes an important part, has been provided numerically. Finally, we show in Fig. 5 that reservoir information is successfully transferred to the stabilized Cat-Qubit using the collision method. This model is particularly significant for the energy-dissipative variant of binary quantum classification, which is the fundamental processing unit of quantum machine learning algorithms. Cat-Qubit architecture, on the other hand, provides an alternate hardware possibility for quantum artificial neural networks due to its rich physics and the potential to implement activation-like functions in artificial neural networks.

## V. APPENDIX

TABLE II
EXPLANATIONS OF MATHEMATICAL TERMS

| Presentation | Explanation |
|---|---|
| $\lvert \cdot \rangle$ | Fock state |
| $\lvert \alpha \rangle$ | Coherent state (Cat state in X basis) |
| $\mathcal{N}^{\pm}$ | Normalization factor for cat qubit bases |
| $\lvert C_\alpha^{\pm} \rangle$ | Cat state in standard basis |
| $\lvert C_\alpha^{i\pm} \rangle$ | Cat state in Y basis |
| $\Phi_t^{(i)}[\varrho_0]$ | Dynamical maps |
| $\Lambda_t[\varrho_0]$ | A weighted combination of quantum dynamical maps |
| $\mathcal{L}[\cdot]$ | Lindblad super operator |

ACKNOWLEDGMENT

The authors thank the funding from TÜBİTAK-Grant No. 120F353. We would also like to thank the Cognitive Systems Laboratory in the Department of Electrical Engineering for providing a suitable environment for the studies.

## BIOGRAPHIES

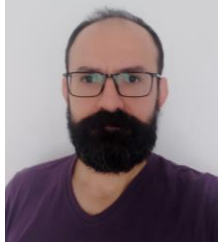

**UFUK KORKMAZ** was born in Turkey. He received the BSc, MSc and PhD degrees from the Ondokuz Mayıs University (OMU), Physics Department, in 2006, 2010 and 2014 respectively. He worked as a Postdoctoral Researcher at Istanbul Technical University (ITU) in 2018-2019. He starting researches as Post-Doc in Istanbul Technical University (ITU) in 2018-2019. His research interests are IR and UV spectroscopy, X-ray single crystal diffraction, Understanding the nature of H bonds in supramolecular structure, Quantum Mechanics and Quantum information theory.

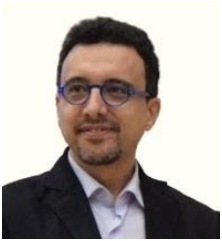

**DENİZ TÜRKPENÇE** Samsun, Turkey in 1977. He completed his primary, secondary and high school education in Samsun. He completed his university education in Samsun Ondokuz Mayıs University, Faculty of Education, Department of Physics in 2000. In 2005, he started his master's degree in Ondokuz Mayıs University, Faculty of Arts and Sciences, Department of Physics. In March 2010, he worked as a researcher at Dortmund Technical University for 1 year with a YÖK scholarship to conduct research and examination abroad related to his doctoral thesis. He received his PhD from Ondokuz Mayıs University in June 2013. He worked as a Research Assistant at Koç University in April 2014 and took part in an international project. He completed her studies at Koç University in 2016. He worked as a postdoctoral researcher in the Cognitive Systems Laboratory of the Electrical Engineering Department of Istanbul Technical University between 2017-2018. He started to work as an Instructor in the Department of Electrical Engineering at Istanbul Technical University in March 2018. In November 2019, he was awarded the title of Associate Professor by the Interuniversity Board. It was decided to support the 3501 TÜBİTAK career project proposed in February 2021. Finally, he is the director of Quantum Systems and Cyber Security Laboratory at ITU Faculty of Electrical and Electronics and he still continues his duty as a Lecturer in ITU Electrical Engineering Department.